\documentclass[12pt]{article}
\usepackage{graphicx}
\usepackage{amsfonts}
\usepackage{indentfirst}
\usepackage{longtable}
\usepackage{epsfig}
\usepackage{axodraw}
\usepackage{cite}

\def\MSbar{$\overline{\mathrm{MS}}\ $}
\def\msbar{\overline{\mathrm{MS}}}

\def\ba{\begin{eqnarray}}
\def\ea{\end{eqnarray}}

\def\dd{{\mathrm d}}

\def\fun#1#2{\lower3.6pt\vbox{\baselineskip0pt\lineskip.9pt
  \ialign{$\mathsurround=0pt#1\hfil##\hfil$\crcr#2\crcr\sim\crcr}}}

\def\gsim{\mathrel{\raise.3ex\hbox{$>$\kern-.75em\lower1ex\hbox{$\sim$}}}}
\def\lsim{\mathrel{\raise.3ex\hbox{$<$\kern-.75em\lower1ex\hbox{$\sim$}}}}

\newcommand{\sss}[1]{\scriptscriptstyle{#1}}
\def\GF {G_{\sss F}}
\def\gw {\Gamma_{\sss W}}
\def\gz {\Gamma_{\sss Z}}
\def\mw {M_{\sss W}}
\def\mz {M_{\sss Z}}
\def\mh {M_{\sss H}}

\newcommand{\GeV}{\unskip\,\mathrm{GeV}}
\newcommand{\MeV}{\unskip\,\mathrm{MeV}}

\title{Inverse bremsstrahlung contributions \\ to Drell--Yan like processes}

\author{A.B. Arbuzov$^{1,2,}$\/\thanks{e-mail: arbuzov@theor.jinr.ru},
\ 
R.R. Sadykov$^2$\ 
}

\date{}

\begin{document}

\maketitle

\begin{center}{\it
$^1$Bogoliubov Laboratory of Theoretical Physics, \\
JINR,\ Dubna, \ 141980 \ \  Russia \\
$^2$Dzhelepov Laboratory of Nuclear Problems, \\
JINR,\ Dubna, \ 141980 \ \  Russia }
\end{center}

\begin{abstract}
The contribution of the sub-process $\gamma q \to q' l_1\bar{l}_2$
in hadron-hadron interactions is considered.  
It is a part of one-loop electroweak radiative corrections for the Drell--Yan
production of lepton pairs at hadron colliders.  
It is shown that this contribution should be taken into account aiming at 
the 1\% accuracy of the Drell--Yan process theoretical description.
Both the neutral and charged current cases are evaluated.
Numerical results are presented for typical conditions of LHC experiments.
\end{abstract}

\section{Introduction}

The Drell--Yan like processes at high energy hadron colliders provide
an advanced tool for precision studies of several problems in the
elementary particle phenomenology. Studies of single $Z$ and $W$
bosons production with the subsequent decays into leptonic pairs play 
a very important role in the physical programs of Tevatron~\cite{Abazov:2003sv,Abachi:1996ey} 
and LHC~\cite{Altarelli:2000ye,Krasnikov:1997sm}.  
These processes have large cross sections and clean signatures in 
the detectors. That allows to reach at LHC the 1\% experimental accuracy
for the total cross sections of these processes as well as high precision
in the measurements of differential distributions.
In particular, Drell--Yan like processes are planned be used at LHC for 
luminosity monitoring, $W$ mass and width measurement, detector calibration, 
extraction of parton density functions, new physics searches, 
and other purposes.

Adequately precise theoretical predictions for single $Z$ and $W$ production at LHC 
are required. For this reason we have to scrutinize several effects involved in the
derivation of the theoretical accuracy: QCD and electroweak radiative corrections,
uncertainties in the partonic density functions (PDF's), technical precision of
Monte Carlo event generators {\it etc.} In this paper we consider a particular
contribution of the first order electroweak radiative corrections coming from
the photon induced process 
\ba \label{process}
h_1\ + h_2\ \to\ X\ + \gamma\ + q\ \to\ X\ +\ q'\ +\ l_1\ +\ \bar{l}_2,
\ea
where $h_{1,2}$ stand for the initial colliding
hadrons; $l_1$ and $\bar{l}_2$ is a pair of leptons ({\it e.g.} $\mu^-$ and $\mu^+$,
or $\nu_e$ and $e^+$); $X+q'$ denotes the remaining final state particles 
(typically they are hadrons). Here $\gamma$ and $q$ are treated as 
partons found in the initial hadrons with certain energy fractions at a given
factorization scale. In this paper we use the MRST2004QED~\cite{Martin:2004dh} 
parameterization of parton density functions (PDFs), which provides in particular the photon 
content in proton at NLO. Note that the evolution~\cite{Roth:2004ti} of the partonic
densities taking into account simultaneous QCD and QED effects leads to the 
unique value of the factorization scale, so that it is impossible to disentangle
QED and QCD contributions. This leads also to the fact that the reduction of the
factorization scale dependence can be reached now only by taking into account 
both QED and QCD higher order radiative corrections and that should be performed
within the same factorization scheme. Nevertheless due to the smallness of the fine 
structure constant $\alpha$ in comparison with the strong coupling constant $\alpha_s$, we can 
limit ourselves to the evaluation of only the first order electroweak 
corrections~\cite{Mosolov:1981xk,Soroko:1990ug,Wackeroth:1996hz,Baur:1998kt,Dittmaier:2001ay,Baur:2001ze,Baur:2004ig,Arbuzov:2005dd,CarloniCalame:2006zq,Zykunov:2006yb} 
together with certain higher order leading logarithmic 
contributions~\cite{CarloniCalame:2003ux,CarloniCalame:2003ck,CarloniCalame:2005vc}. 
At the same moment QCD corrections
have to be treated at least at NNLO~\cite{Anastasiou:2003yy,Anastasiou:2003ds,Melnikov:2006kv}.
Some numerical results for the inverse bremsstrahlung contribution to the charged current case 
(single $W$ boson production) were already presented by S.~Dittmaier and M.~Kr\"amer in 
the proceedings 
of the Les Houches workshop~\cite{Buttar:2006zd}. We performed an independent calculation and give
below a comparison with the earlier results. The neutral current case is considered in addition.

This paper is organized as follows. In the next section we present the derivation
of the Drell--Yan process cross sections in the scheme with massive quarks. The subtraction
of the quark mass singularities is described in Sect.~\ref{SUBTR}. Numerical results and their
discussion are presented in Conclusions.

\section{Inverse bremsstrahlung with massive quarks}
  
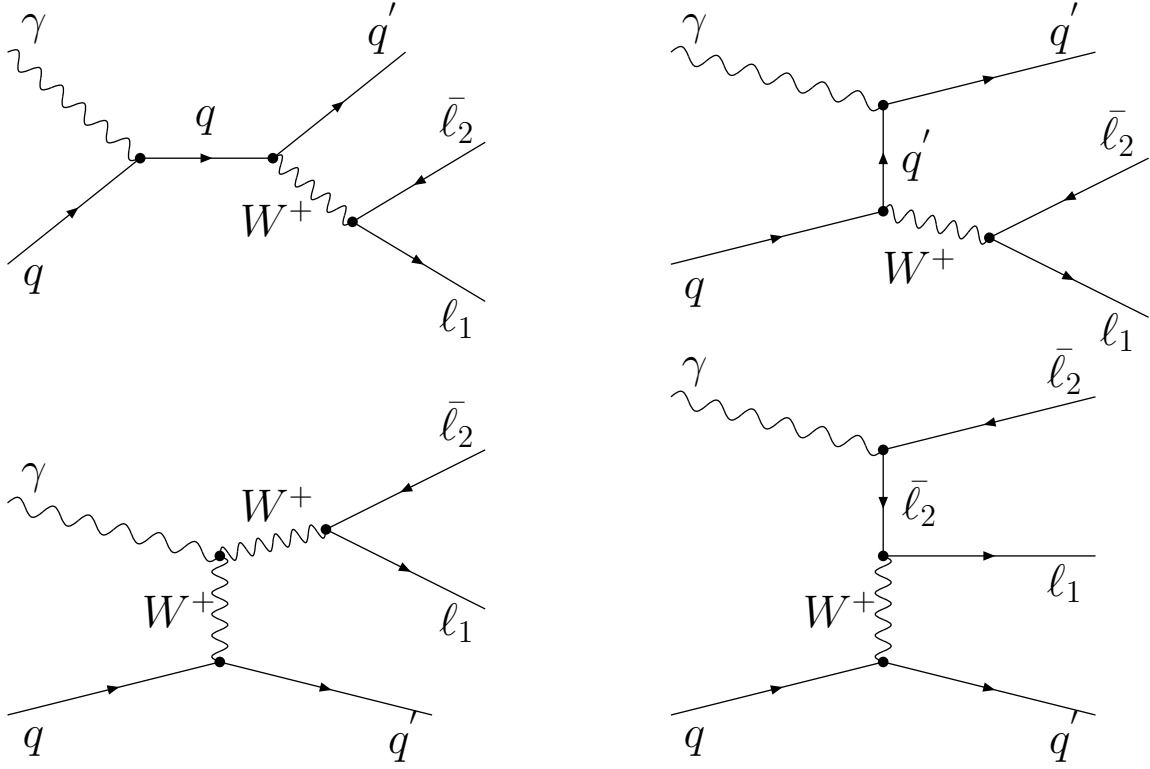
\begin{figure}[h]
\[
\begin{picture}(480,280)(0,0)
  \ArrowLine(20,20)(100,40)
  \Vertex(100,40){2}
  \ArrowLine(100,40)(180,20)
  \Photon(100,40)(100,80){3}{5}
  \Photon(20,100)(100,80){3}{6}
  \Vertex(100,80){2}
  \Photon(100,80)(140,90){3}{6}
  \Vertex(140,90){2}
  \ArrowLine(200,120)(140,90)
  \ArrowLine(140,90)(200,60)
\Text(30,110)[]{\Large $\gamma$}
\Text(30,10)[]{\Large $q$}
\Text(170,12)[]{\Large $q^{'}$}
\Text(190,55)[]{\Large $\ell_1$}
\Text(190,130)[]{\Large $\bar{\ell_2}$}
\Text(122,100)[]{\Large $W^+$}
\Text(85,60)[]{\Large $W^+$}
  \ArrowLine(270,20)(350,40)
  \Vertex(350,40){2}
  \ArrowLine(350,40)(430,20)
  \Photon(350,40)(350,80){3}{5}
  \Vertex(350,80){2}
  \ArrowLine(350,120)(350,80)
  \ArrowLine(350,80)(430,80)
  \Photon(270,140)(350,120){3}{6}
  \Vertex(350,120){2}
  \ArrowLine(430,140)(350,120)
\Text(280,150)[]{\Large $\gamma$}
\Text(280,10)[]{\Large $q$}
\Text(420,12)[]{\Large $q^{'}$}
\Text(420,70)[]{\Large $\ell_1$}
\Text(420,150)[]{\Large $\bar{\ell_2}$}
\Text(365,100)[]{\Large $\bar{\ell_2}$}
\Text(335,60)[]{\Large $W^+$}
  \Photon(20,270)(70,230){3}{6}
  \ArrowLine(20,190)(70,230)
  \Vertex(70,230){2}
  \ArrowLine(70,230)(120,230)
  \Vertex(120,230){2}
  \ArrowLine(120,230)(170,270)
  \Photon(120,230)(150,206){3}{5}
  \Vertex(150,206){2}
  \ArrowLine(200,236)(150,206)
  \ArrowLine(150,206)(200,176)
\Text(30,280)[]{\Large $\gamma$}
\Text(30,185)[]{\Large $q$}
\Text(162,280)[]{\Large $q^{'}$}
\Text(190,170)[]{\Large $\ell_1$}
\Text(190,245)[]{\Large $\bar{\ell_2}$}
\Text(95,245)[]{\Large $q$}
\Text(122,208)[]{\Large $W^+$}
  \ArrowLine(270,190)(350,210)
  \Vertex(350,210){2}
  \Photon(350,210)(390,200){3}{5}
  \ArrowLine(350,210)(350,250)
  \Photon(270,270)(350,250){3}{6}
  \Vertex(350,250){2}
  \ArrowLine(350,250)(430,270)
  \Vertex(390,200){2}
  \ArrowLine(450,230)(390,200)
  \ArrowLine(390,200)(450,170)
\Text(280,280)[]{\Large $\gamma$}
\Text(280,180)[]{\Large $q$}
\Text(420,280)[]{\Large $q^{'}$}
\Text(440,165)[]{\Large $\ell_1$}
\Text(440,240)[]{\Large $\bar{\ell_2}$}
\Text(364,232)[]{\Large $q^{'}$}
\Text(365,190)[]{\Large $W^+$}
\end{picture}
\]
\vspace*{-6mm}
\caption{Feynman diagrams for inverse bremsstrahlung in the charged current Drell--Yan sub-process.}
\label{Feyn_CC}
\end{figure}

\begin{figure}[h]
\[
\begin{picture}(480,269)(0,0)
  \ArrowLine(20,20)(100,40)
  \Vertex(100,40){2}
  \ArrowLine(100,40)(180,20)
  \Photon(100,40)(100,80){3}{5}
  \Vertex(100,80){2}
  \ArrowLine(100,80)(100,120)
  \ArrowLine(180,80)(100,80)
  \Photon(20,140)(100,120){3}{6}
  \Vertex(100,120){2}
  \ArrowLine(100,120)(180,140)
\Text(30,150)[]{\Large $\gamma$}
\Text(30,10)[]{\Large $q$}
\Text(170,12)[]{\Large $q$}
\Text(170,70)[]{\Large $\bar{\ell_1}$}
\Text(170,150)[]{\Large $\ell_1$}
\Text(115,100)[]{\Large $\ell_1$}
\Text(82,60)[]{\Large $\gamma,Z$}
  \ArrowLine(270,20)(350,40)
  \Vertex(350,40){2}
  \ArrowLine(350,40)(430,20)
  \Photon(350,40)(350,80){3}{5}
  \Vertex(350,80){2}
  \ArrowLine(350,120)(350,80)
  \ArrowLine(350,80)(430,80)
  \Photon(270,140)(350,120){3}{6}
  \Vertex(350,120){2}
  \ArrowLine(430,140)(350,120)
\Text(280,150)[]{\Large $\gamma$}
\Text(280,10)[]{\Large $q$}
\Text(420,12)[]{\Large $q$}
\Text(420,70)[]{\Large $\ell_1$}
\Text(420,150)[]{\Large $\bar{\ell_1}$}
\Text(365,100)[]{\Large $\bar{\ell_1}$}
\Text(332,60)[]{\Large $\gamma,Z$}
  \Photon(20,270)(70,230){3}{6}
  \ArrowLine(20,190)(70,230)
  \Vertex(70,230){2}
  \ArrowLine(70,230)(120,230)
  \Vertex(120,230){2}
  \ArrowLine(120,230)(170,270)
  \Photon(120,230)(150,206){3}{5}
  \Vertex(150,206){2}
  \ArrowLine(200,236)(150,206)
  \ArrowLine(150,206)(200,176)
\Text(30,280)[]{\Large $\gamma$}
\Text(30,185)[]{\Large $q$}
\Text(162,280)[]{\Large $q$}
\Text(190,170)[]{\Large $\ell_1$}
\Text(190,245)[]{\Large $\bar{\ell_1}$}
\Text(95,245)[]{\Large $q$}
\Text(120,208)[]{\Large $\gamma,Z$}
  \ArrowLine(270,190)(350,210)
  \Vertex(350,210){2}
  \Photon(350,210)(390,200){3}{5}
  \ArrowLine(350,210)(350,250)
  \Photon(270,270)(350,250){3}{6}
  \Vertex(350,250){2}
  \ArrowLine(350,250)(430,270)
  \Vertex(390,200){2}
  \ArrowLine(450,230)(390,200)
  \ArrowLine(390,200)(450,170)
\Text(280,280)[]{\Large $\gamma$}
\Text(280,180)[]{\Large $q$}
\Text(420,280)[]{\Large $q$}
\Text(440,165)[]{\Large $\ell_1$}
\Text(440,240)[]{\Large $\bar{\ell_1}$}
\Text(362,230)[]{\Large $q$}
\Text(365,190)[]{\Large $\gamma,Z$}
\end{picture}
\]
\vspace*{-6mm}
\caption{Feynman diagrams for inverse bremsstrahlung in the neutral current Drell--Yan sub-process.}
\label{Feyn_NC}
\end{figure}
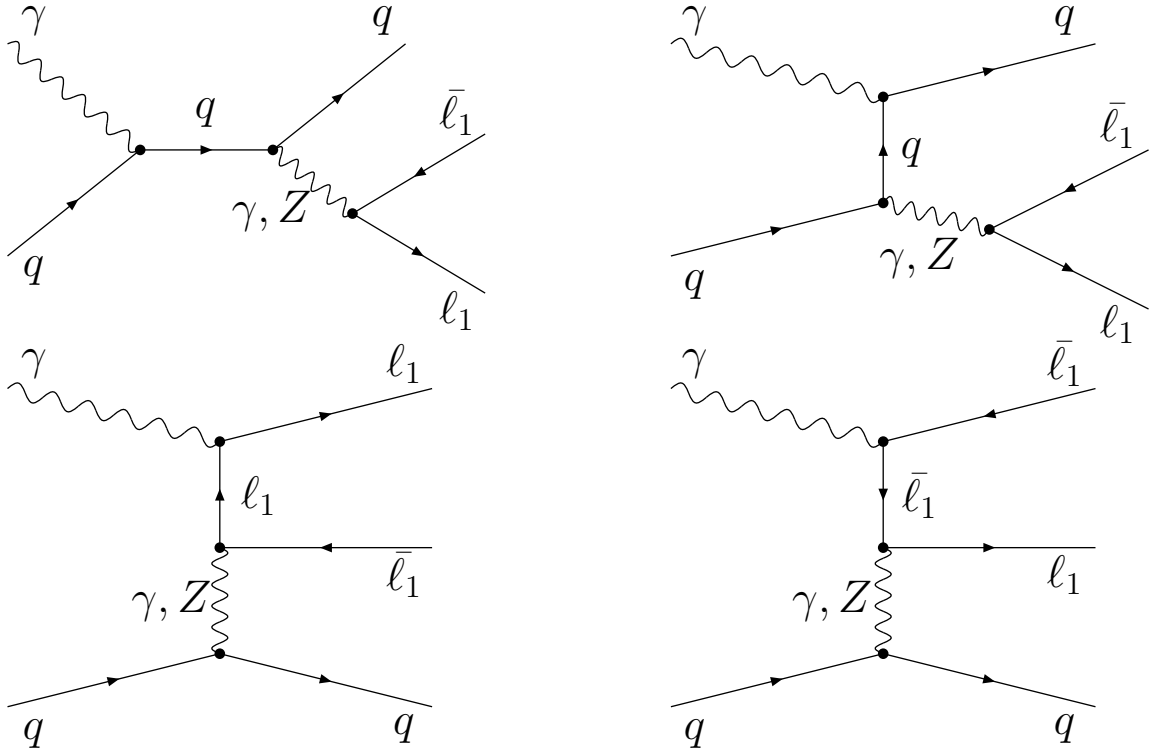

Let us compute the cross section of the process~(\ref{process}) in the form proposed by 
Drell and Yan~\cite{Drell:1970wh} as of a convolution of the parton density functions 
with the hard sub-process distribution. In our case the sub-processes is
\ba
q\ +\ \gamma\ \to\ q'\ +\ l_1\ +\ \bar{l}_2,
\ea
where quarks $q$ and $q'$ are of the same type for the neutral current (NC) case
and different for the charged current (CC) one. 
We compute the matrix element of the NC and CC sub-processes with help of the
SANC system~\cite{Andonov:2004hi,Bardin:2005dp} environment. In the actual calculation, 
we start within the massive quark scheme. The matching of this scheme with the
PDF formalism will be performed by means of the subsequent subtraction of
the quark mass singularities from the computed cross section. 
So, we evaluate the
complete tree--level matrix elements of the sub-processes in the standard way keeping
the exact dependence on the quark and lepton masses. The Feynman diagrams 
for the sub-processes under consideration are shown in Figs.~\ref{Feyn_CC} and
\ref{Feyn_NC}.


We construct the squares of the matrix elements in the usual way and obtain the partonic cross 
sections of the sub-processes. These quantities have to be convoluted then with the parton 
density functions:
\ba \label{xs_hadronic}
\frac{\dd\sigma_{\mathrm{inv.brem.}}^{pp\to l_1\bar{l}_2X}(s)}{\dd c_1} 
= \sum\limits_{q_i}\int\limits_{0}^{1}\!\! \int\limits_{0}^{1}\!\!
\dd x_1 \dd x_2 q_i(x_1,M^2) 
\gamma(x_2,M^2)
\frac{\dd^2\hat{\sigma}^{q_i\gamma\to q_i' l_1\bar{l}_2}(\hat{s})}
{\dd\hat{c}_1}{\mathcal J}\Theta(c_1,x_1,x_2),
\ea
where $c_1$ denotes the cosine of the scattering angle of the first lepton
(another variable can be chosen as well).
The step function $\Theta(c_1,x_1,x_2)$ defines the phase space domain 
corresponding to the given event selection procedure.
The partonic cross section is taken in the center-of-mass reference 
frame of the initial partons, where the cosine of the first lepton scattering
angle, $\hat{c}_1$, is defined.
The transformation into the observable variable $c_1$  
involves the Jacobian:
\ba \label{Jacobian}
&& {\mathcal J} = \frac{\partial \hat{c}}{\partial c}  = \frac{4x_1x_2}{a^2}\, ,\quad
a = x_1 + x_2 - c(x_1 - x_2), 
\nonumber \\
&& \hat{c} = 1 - (1-c)\frac{2x_1}{a}, \qquad
\hat{s} = sx_1x_2,
\ea
where $s$ is the squared center-of-mass energy of the colliding hadrons.
An analogous formula can be written for any other choice of a differential
distribution as well as for the total cross section.

In Eq.~(\ref{xs_hadronic}) we presented the contribution, when the photon is
found in the first of the colliding hadrons and the quark is taken from the other
one. Of course, there is also the contribution when we choose the particles in
the other way round, and it is taken into account in our numerical simulations.


\section{Subtraction of the quark mass singularities}  \label{SUBTR}

Since the calculation of the partonic cross sections was performed keeping finite
masses of the quarks, the result~(\ref{xs_hadronic}) depends on the values of the masses.
For the high energies this dependence arises in the form of large logarithms
of the type $\ln(M^2/m_q^2)$ that give a considerable numerical effect,
while the other mass-dependent contributions suppressed by the factor $m^2_q/M^2\ll 1$
can be omitted (here $M$ is a typical energy scale of the partonic sub-process).
The large logarithms represent quark mass singularities. They  
can be treated with help of the QED renormalization group approach.
But the point is that they have been already taken into account in
the evolution of partonic density functions. In the MRST2004QED distributions~\cite{Martin:2004dh}
that have been done explicitly. But even in any other PDF on the market the QED evolution
is implicitly taken into account since it has not been subtracted from the experimental data before 
the PDF fitting procedure. In fact, QED corrections to the quark line in deep inelastic scattering 
are usually omitted in the data analysis, see Refs.~\cite{Akhundov:1994my,Arbuzov:1995id}.

The quark mass singularity of the first type arises from the right Feynman diagrams in the 
upper line of 
Fig.~\ref{Feyn_CC} and of Fig.~\ref{Feyn_NC} for CC and NC cases, respectively. 
The singularity originates from the kinematical domain when the virtual quark propagator 
is close to the mass shell. For this situation there is a convolution of distributions 
of the two sub-processes:
a conversion of the photon into a pair of quarks and a Drell--Yan partonic process 
$q'+q\to l_1\bar{l}_2$. In the \MSbar scheme the corresponding contribution reads
\ba \label{delta1}
\delta_1(c_1) &=&
\sum\limits_{q_i}\int\limits_{0}^{1} \int\limits_{0}^{1} 
\dd x_1\; \dd x_2\; \gamma(x_1,M^2) q_i(x_2,M^2) 
\int\limits_{0}^{1}\dd x_3 D_{q'\gamma}(x_3,M,m_{q'})
\nonumber \\
&\times& \frac{\dd^2\tilde{\sigma}^{q_i q_i'\to l_1\bar{l}_2}(\tilde{s})}
{\dd\tilde{c}_1}\;\tilde{\mathcal J}\;\Theta(c_1,x_1x_3,x_2),
\ea
where $\tilde{c}_1$, $\tilde{\mathcal J}$  and $\tilde{s}$ are calculated
according to Eq.~(\ref{Jacobian}) with the interchange $x_1\to x_1x_3$. 
For the NC case in the above equation we have $q'=q$. 
The structure function $D_{q'\gamma}(x_3,M,m_{q'})$ describes the probability to find
quark $q'$ with energy fraction $x_3$ in the photon. For the \MSbar 
scheme at NLO this function reads
\ba
&& D_{q'\gamma}^{\msbar}(x_3,M,m_{q'}) = \frac{\alpha}{2\pi}Q_{q'}^2\ln\frac{M^2}{m_{q'}^2}
[x_3^2+(1-x_3)^2],
\ea
where $M$ is the factorization scale, and $Q_{q'}$ is the quark charge. 

In the neutral current case there is one additional source of the quark mass
singularities. It arises from the two lower Feynman amplitudes in Fig.~\ref{Feyn_NC},
when the virtual photon propagator is near the mass shell. In this case we have
the convolution of the distributions of the following processes: $2\gamma\to l_1\bar{l}_1$
and $q\to \gamma q$. The corresponding contribution is
\ba\label{delta2}
\delta_2(c_1) &=&  
\sum\limits_{q_i}\int\limits_{0}^{1} \int\limits_{0}^{1} 
\dd x_1\; \dd x_2\; q_i(x_1,M^2) 
\gamma(x_2,M^2) \int\limits_{0}^{1}\dd x_3 D_{\gamma q}(x_3,M,m_{q})
\nonumber \\
&\times& \frac{\dd^2\tilde{\sigma}^{\gamma\gamma\to l_1\bar{l}_1}(\tilde{s})}
{\dd\tilde{c}_1}\;\tilde{\mathcal J}\;\Theta(c_1,x_1x_3,x_2).
\ea
The relevant structure function describes the probability to find a photon
with a certain energy fraction in the quark:
\ba
&& D_{\gamma q}^{\msbar}(x_3,M,m_{q}) = \frac{\alpha}{2\pi}Q_{q'}^2
\frac{1+(1-x_3)^2}{x_3}\biggl\{\ln\frac{M^2}{m_{q}^2} - 2\ln x_3 - 1 \biggr\},
\ea 

According to the renormalization formalism we have now to subtract the
contributions (\ref{delta1}) and (\ref{delta2}) from the computed cross section~(\ref{xs_hadronic}).
In a realistic situation we have to perform this procedure numerically in order to keep
the possibility to impose experimental cuts. On the other hand, it can be shown analytically
that the terms with the logarithms of the quark masses do cancel out during the subtraction
procedure.

\section{Numerical Results and Conclusions}
\setcounter{table}{0}

For the numerical evaluations we used
the same conditions and the input parameters as
in Ref.~\cite{Buttar:2006zd}:
\ba \nonumber
\begin{array}[b]{lcllcllcl}
G_F & = & 1.16637 \times 10^{-5} \GeV^{-2}, & && \\
\alpha(0) &=& 1/137.03599911, & 
\alpha_s &=& 0.1187, \\
\mw & = & 80.425\GeV, &
\gw & = & 2.124\GeV, \\
\mz & = & 91.1867\GeV,& 
\gz & = & 2.4952\GeV, \\
\mh & = & 150\GeV, &
m_t & = & 174.17\;\GeV, \\
m_u & = & m_d = 66\;\MeV, &
m_c & = & 1.55\;\GeV, \\
m_s & = & 150\;\MeV, &
m_b & = & 4.5\;\GeV, \\
|V_{ud}| & = & |V_{cs}| = 0.975, &
|V_{us}| & = & |V_{cd}| = 0.222. 
\end{array}
\ea
The MRST204QED set~\cite{Martin:2004dh} of PDF's and the $\GF$ EW scheme 
were used. The cut on the charged lepton rapidity and transverse momentum
are $|\eta_\ell|< 1.2$ and $P_{T,\ell}>25$~GeV.
The cut on the missing transverse momentum for the CC case is imposed as well:
$P_{T,\mathrm{missing}}>25$~GeV.

At the partonic level for the CC and NC processes ($\gamma + q \to q' + l_1 + \bar{l}_2$)
we performed a comparison with the corresponding distributions obtained with
help of the CompHEP system~\cite{Boos:2004kh} and found a good agreement.

\begin{table}[h]
{\small\small
\begin{tabular}{|l|l|l|l|l|l|l|}
\hline
$P_{T,\mu}/\mathrm{GeV}$&
$\mathrm{25-\infty}$&
$\mathrm{50-\infty}$&
$\mathrm{100-\infty}$&
$\mathrm{200-\infty}$&
$\mathrm{500-\infty}$&
$\mathrm{1000-\infty}$\\
\hline
\multicolumn{7} {|l|} {$\mathrm{\sigma_0/pb}$}\\
\hline
$\mathrm{ DK}$&
$\mathrm{ 2112.2(1)}$&
$\mathrm{ 13.152(2)}$&
$\mathrm{ 0.9452(1)}$&
$\mathrm{ 0.11511(2)}$&
$\mathrm{ 0.0054816(3)}$&
$\mathrm{ 0.00026212(1)}$\\
\hline
$\mathrm{ SANC}$&
$\mathrm{ 2112.2(1)}$&
$\mathrm{ 13.151(1)}$&
$\mathrm{ 0.9451(1)}$&
$\mathrm{ 0.11511(1)}$&
$\mathrm{ 0.0054813(1)}$&
$\mathrm{ 0.00026211(1)}$\\
\hline
\multicolumn{7} {|l|} {$\mathrm{\delta_{\gamma q}/\%}$}\\
\hline
$\mathrm{ DK}$&
$\mathrm{ 0.071(1)}$&
$\mathrm{ 5.24(1)}$&
$\mathrm{ 13.10(1)}$&
$\mathrm{ 16.44(2)}$&
$\mathrm{ 14.30(1)}$&
$\mathrm{ 11.89(1)}$\\
\hline
$\mathrm{ SANC}$&
$\mathrm{ 0.074(1)}$&
$\mathrm{ 5.24(1)}$&
$\mathrm{ 13.09(1)}$&
$\mathrm{ 16.43(1)}$&
$\mathrm{ 14.30(1)}$&
$\mathrm{ 11.90(1)}$\\
\hline
\end{tabular}
}
\caption{Cross sections $\mathrm{\sigma_{0}}$ and $\mathrm{\sigma_{\gamma q}}$
of the processes $\mathrm{p[q]p[q'] \to \nu_\mu \mu^+ X}$ and
$\mathrm{p[\gamma]p[q] \to \nu_\mu \mu^+ X}$, respectively
and corresponding corrections $\mathrm{\delta_{\gamma q} = \sigma_{\gamma q}/\sigma_{0}}$,
obtained by DK and SANC groups for different $\mathrm{P_{T,\mu}}$ ranges at LHC.}
\label{table1}
\end{table}
In Table~\ref{table1} we present the results of comparison for the inverse bremsstrahlung
contribution to the CC Drell--Yan process with different cuts on the charged lepton transverse 
momentum (see the details in Ref.~\cite{Buttar:2006zd}). 
Our results are marked as ``SANC'', they are compared with the numbers (``DK'') presented
by the S.~Dittmaier and M.~Kr\"amer in Ref.~\cite{Buttar:2006zd}. The small deviations
in the results for the values of the corrections are certainly beyond the 1\% precision level.
They are due to some differences in the schemes
of calculations and are induced by higher order effects in $\alpha$.

\begin{table}[h]
{\small
\begin{tabular}{|l|l|l|l|l|l|l|}
\hline
$\mathrm{M_{T,\nu_\mu\mu^+}/GeV}$&
$\mathrm{50-\infty}$&
$\mathrm{100-\infty}$&
$\mathrm{200-\infty}$&
$\mathrm{500-\infty}$&
$\mathrm{1000-\infty}$&
$\mathrm{2000-\infty}$\\
\hline
\multicolumn{7} {|l|} {$\mathrm{\sigma_0/pb}$}\\
\hline
$\mathrm{ DK}$&
$\mathrm{ 2112.2(1)}$&
$\mathrm{ 13.152(2)}$&
$\mathrm{ 0.9452(1)}$&
$\mathrm{ 0.057730(5)}$&
$\mathrm{ 0.0054816(3)}$&
$\mathrm{ 0.00026212(1)}$\\
\hline
$\mathrm{ SANC}$&
$\mathrm{ 2112.2(1)}$&
$\mathrm{ 13.151(1)}$&
$\mathrm{ 0.9451(1)}$&
$\mathrm{ 0.057730(5)}$&
$\mathrm{ 0.0054813(1)}$&
$\mathrm{ 0.00026211(1)}$\\
\hline
\multicolumn{7} {|l|} {$\mathrm{\delta_{\gamma q}/\%}$}\\
\hline
$\mathrm{ DK}$&
$\mathrm{  0.0567(3)}$&
$\mathrm{  0.1347(1)}$&
$\mathrm{  0.2546(1)}$&
$\mathrm{  0.3333(1)}$&
$\mathrm{  0.3267(1)}$&
$\mathrm{  0.3126(1)}$\\
\hline
$\mathrm{ SANC}$&
$\mathrm{  0.0532(1)}$&
$\mathrm{  0.1350(1)}$&
$\mathrm{  0.2537(1)}$&
$\mathrm{  0.3314(1)}$&
$\mathrm{  0.3245(1)}$&
$\mathrm{  0.3094(1)}$\\
\hline
\end{tabular}
}
\caption{Cross sections $\mathrm{\sigma_{0}}$ and $\mathrm{\sigma_{\gamma q}}$
of the processes $\mathrm{p[q]p[q'] \to \nu_\mu \mu^+ X}$ and
$\mathrm{p[\gamma]p[q] \to \nu_\mu \mu^+ X}$, respectively
and corresponding corrections $\mathrm{\delta_{\gamma q} = \sigma_{\gamma q}/\sigma_{0}}$,
obtained by DK and SANC groups for different $\mathrm{M_{T,\nu_\mu\mu^+}}$ ranges at LHC.}
\label{table2}
\end{table}
Table~\ref{table2} shows the results of comparison for the inverse bremsstrahlung
contribution to CC Drell--Yan process with different cuts on transverse mass of muon-neutrino
pair. The corresponding numbers for $\delta_{\gamma q}$ are below percent level.

\begin{table}[h]
{\small
\begin{tabular}{|l|l|l|l|l|l|l|}
\hline
$\mathrm{M_{\mu^+\mu^-}/GeV}$&
$\mathrm{50-\infty}$&
$\mathrm{100-\infty}$&
$\mathrm{200-\infty}$&
$\mathrm{500-\infty}$&
$\mathrm{1000-\infty}$&
$\mathrm{2000-\infty}$\\
\hline
\multicolumn{7} {|l|} {$\mathrm{\sigma_0/pb}$}\\
\hline
$\mathrm{ HORACE}$&
$\mathrm{ 254.64(1)}$&
$\mathrm{ 10.571(1)}$&
$\mathrm{ 0.45303(3)}$&
$\mathrm{ 0.026996(2)}$&
$\mathrm{ 0.0027130(2)}$&
$\mathrm{ 0.00015525(1)}$\\
\hline
$\mathrm{ SANC}$&
$\mathrm{ 254.65(2)}$&
$\mathrm{ 10.571(1)}$&
$\mathrm{ 0.45308(3)}$&
$\mathrm{ 0.026996(2)}$&
$\mathrm{ 0.0027131(2)}$&
$\mathrm{ 0.00015525(1)}$\\
\hline
\multicolumn{7} {|l|} {$\mathrm{\delta_{\gamma q}/\%}$}\\
\hline
\hline
$\mathrm{ SANC}$&
$\mathrm{  0.047(1)}$&
$\mathrm{  0.449(1)}$&
$\mathrm{  0.013(1)}$&
$\mathrm{  0.496(1)}$&
$\mathrm{  0.619(1)}$&
$\mathrm{  0.563(1)}$\\
\hline
\end{tabular}
}
\caption{
Cross sections $\mathrm{\sigma_{0}}$ and $\mathrm{\sigma_{\gamma q}}$
of the processes $\mathrm{p[q]p[q'] \to \mu^+ \mu^- X}$ and
$\mathrm{p[\gamma]p[q] \to \mu^+ \mu^- X}$, respectively
and corresponding corrections $\mathrm{\delta_{\gamma q} = \sigma_{\gamma q}/\sigma_{0}}$,
for different $\mathrm{M_{\mu^+\mu^-}}$ ranges at LHC.}
\label{table3}
\end{table}
Table~\ref{table3} gives the results for the inverse bremsstrahlung contribution
to the neutral current Drell--Yan process with production of two muons. Different values
of the cut on the invariant mass of the muon pair are considered. For the Born cross section
we show also the numbers of HORACE~\cite{CarloniCalame:2003ux,CarloniCalame:2005vc}, 
which are in fair agreement with the SANC results.

\begin{figure}[h]
\begin{center}
\includegraphics*[width=7.9cm,height=7.6cm,keepaspectratio,angle=0]{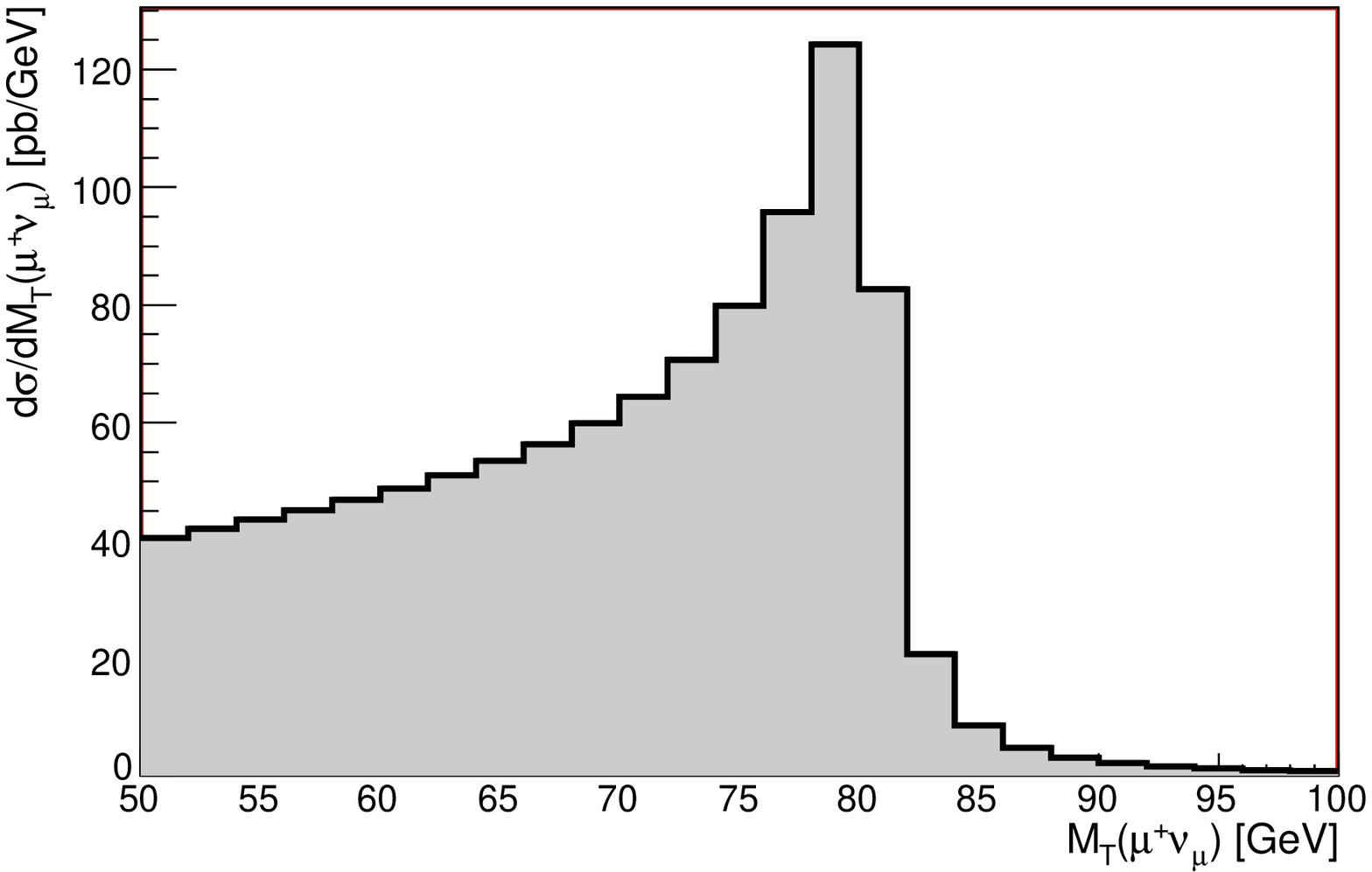}
\includegraphics*[width=7.9cm,height=7.6cm,keepaspectratio,angle=0]{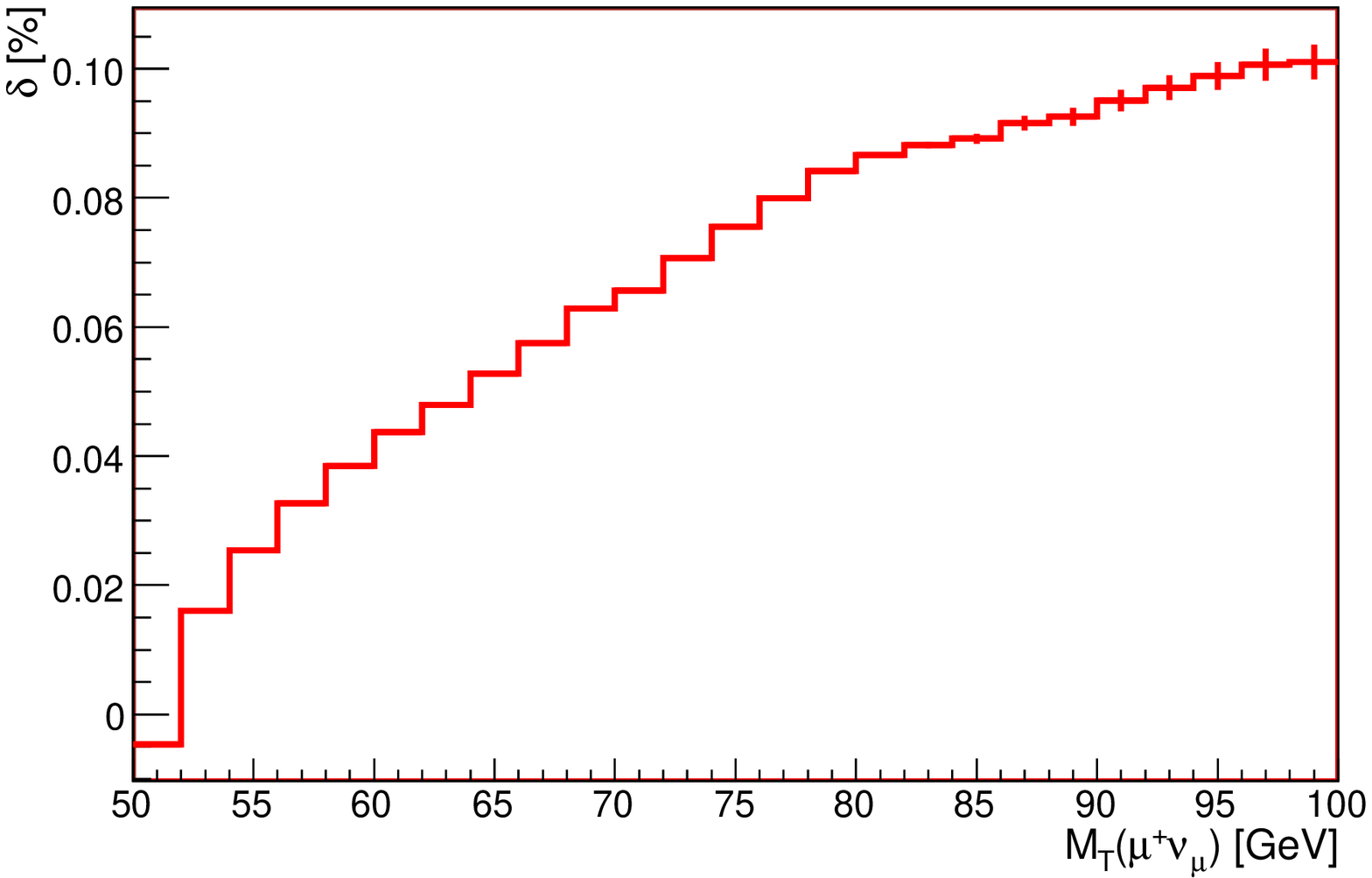} \\
{\bf (a)} \hspace{7cm} {\bf (b)}
\end{center}
\caption{The Born-level CC Drell--Yan cross section and the relative contribution 
of the inverse bremsstrahlung versus the transverse mass of the muon-neutrino pair.
}
\label{fig3}
\end{figure}

\begin{figure}[h]
\begin{center}
\includegraphics*[width=7.9cm,height=7.6cm,keepaspectratio,angle=0]{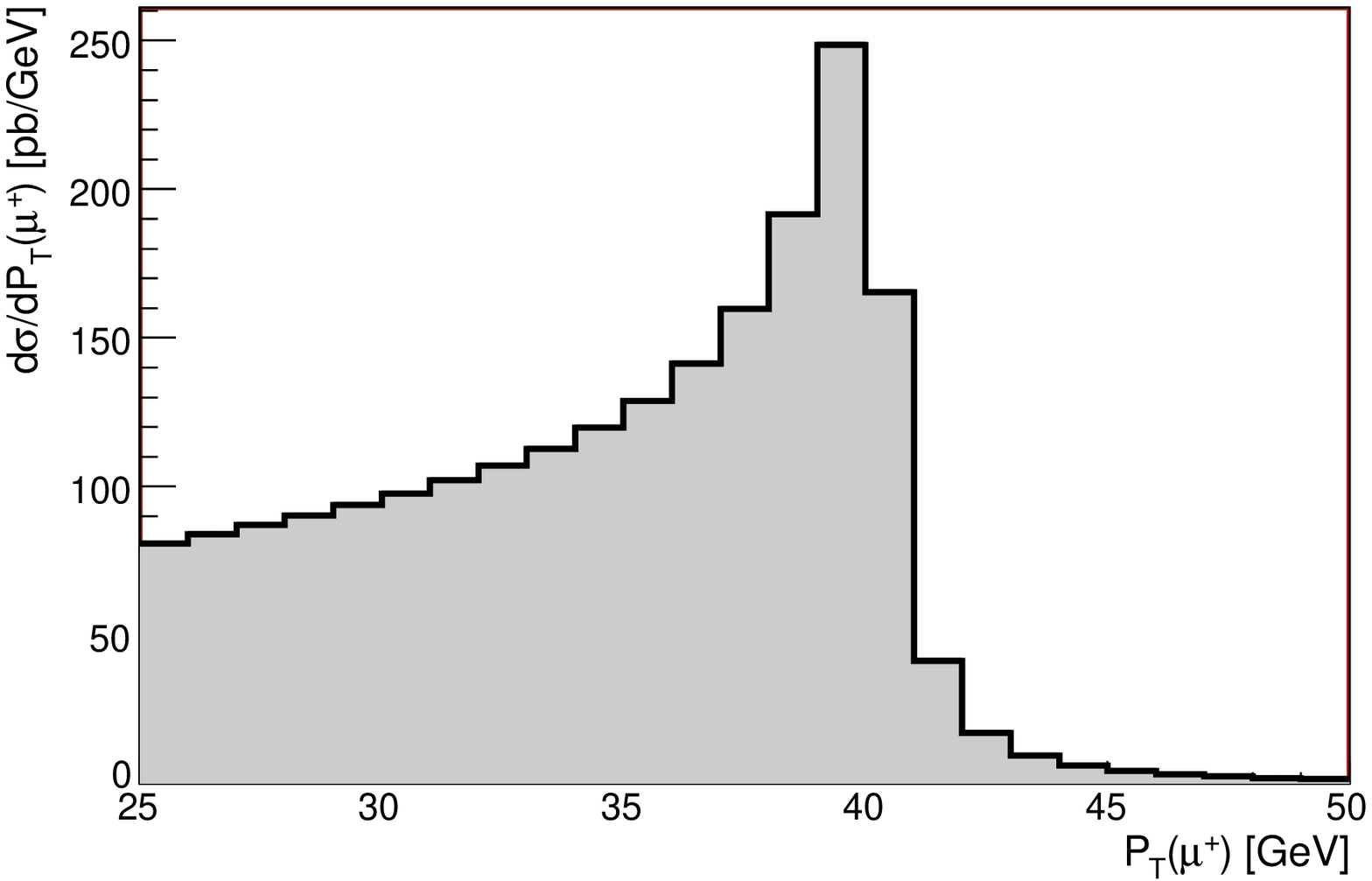}
\includegraphics*[width=7.9cm,height=7.6cm,keepaspectratio,angle=0]{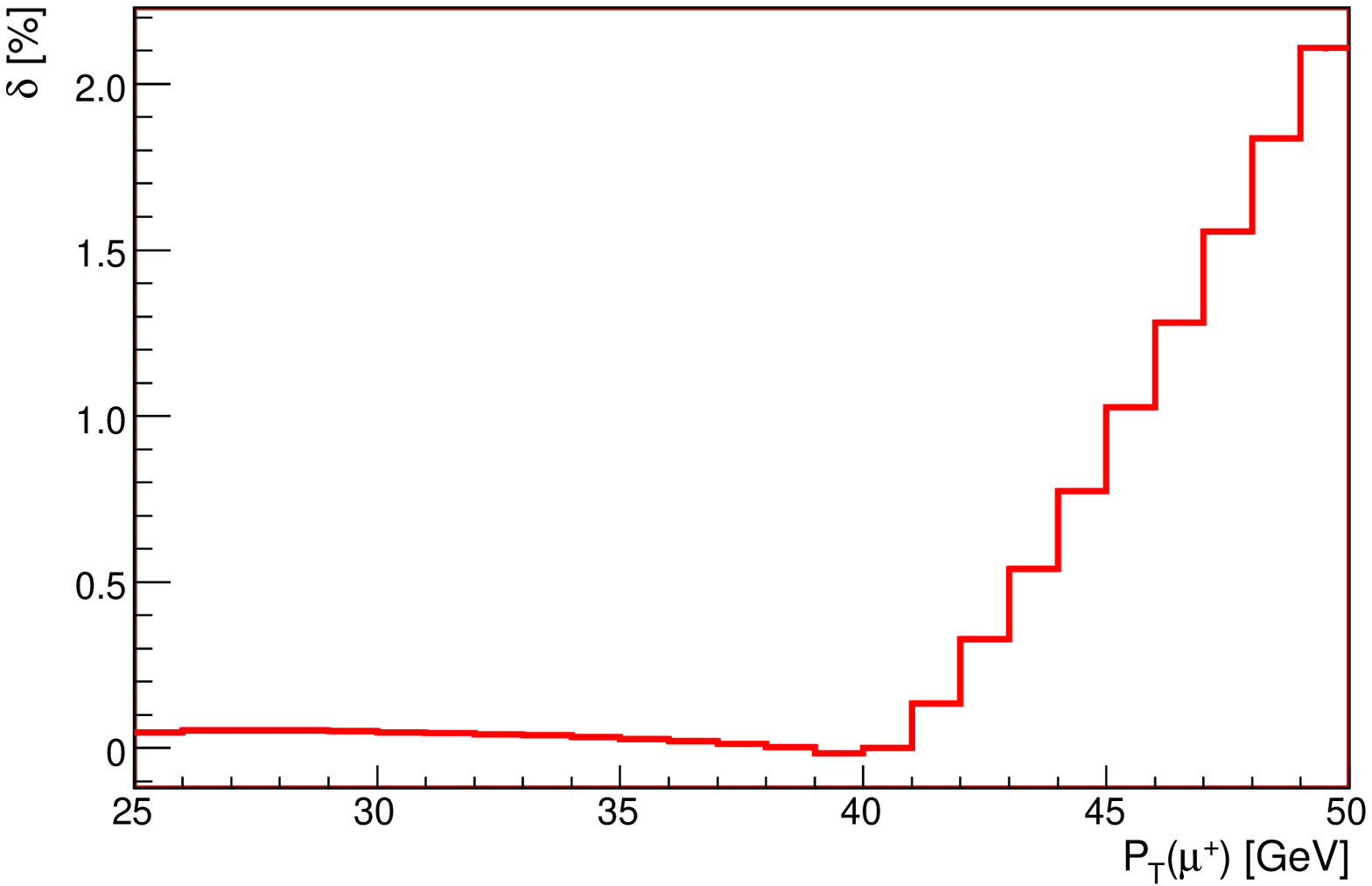} \\
{\bf (a)} \hspace{7cm} {\bf (b)}
\end{center}
\caption{The Born-level CC Drell--Yan cross section and the relative contribution 
of the inverse bremsstrahlung versus the $\mu^+$ transverse momentum.
}
\label{fig4}
\end{figure}
In Fig.~\ref{fig3} we plotted the distributions of the Born-level cross section
and of the relative radiative correction versus the transverse mass of the muon and neutrino
pair $M_T(\mu^+\nu_\mu)$ in the CC Drell-Yan process,
\ba
M_T(\mu^+\nu_\mu) = \sqrt{2P_{T,\mu}P_{T,\nu}(1-\cos\phi_{\mu\nu})},
\ea
where $\phi_{\mu\nu}$ is the angle between the muon momentum and the missing one in 
the transverse plane. In Fig.~\ref{fig4} the analogous distributions in the muon transverse 
momentum $P_{T,\mu}$ are given.

\begin{figure}[h]
\begin{center}
\includegraphics*[width=7.9cm,height=7.6cm,keepaspectratio,angle=0]{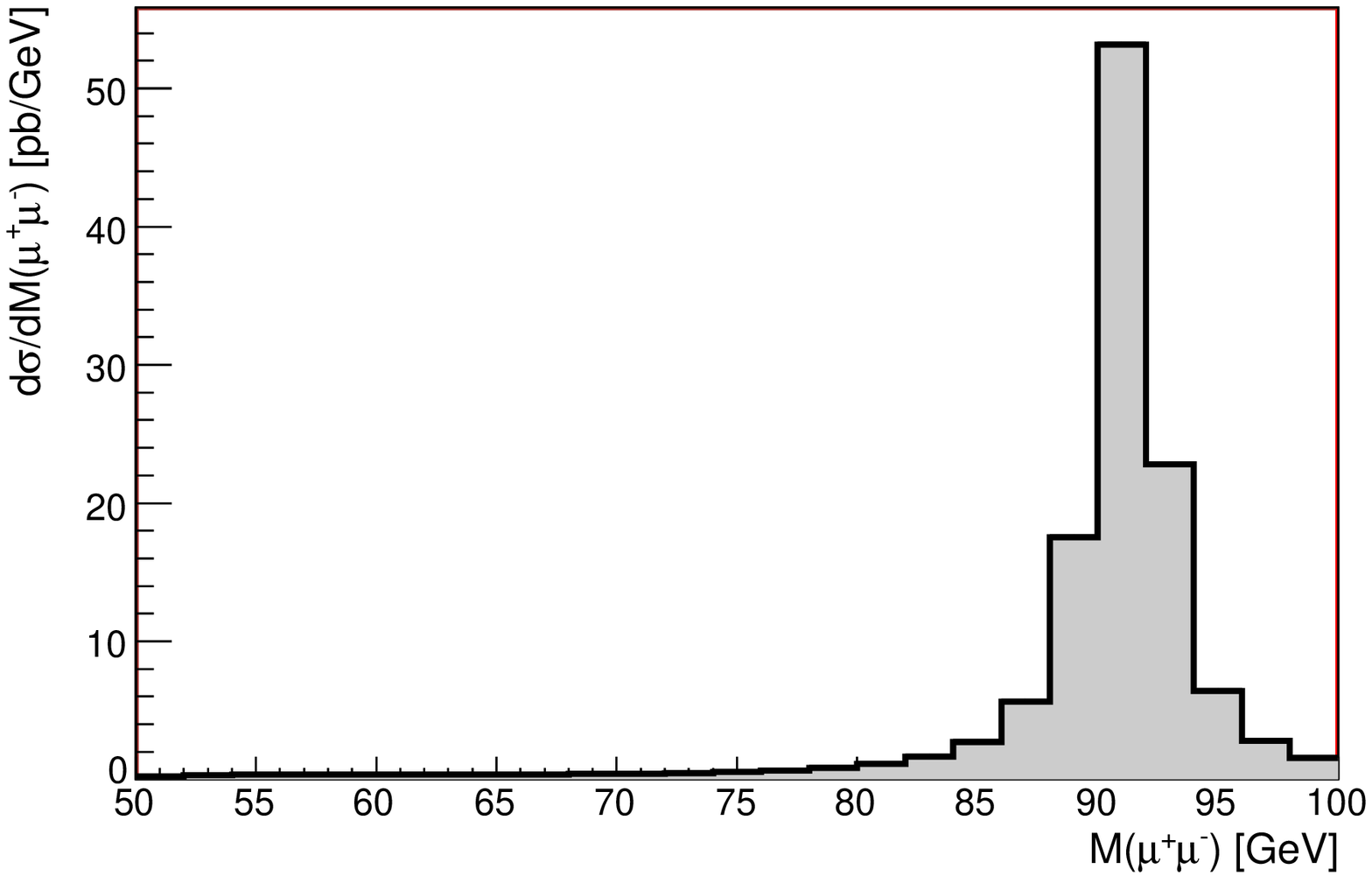}
\includegraphics*[width=7.9cm,height=7.6cm,keepaspectratio,angle=0]{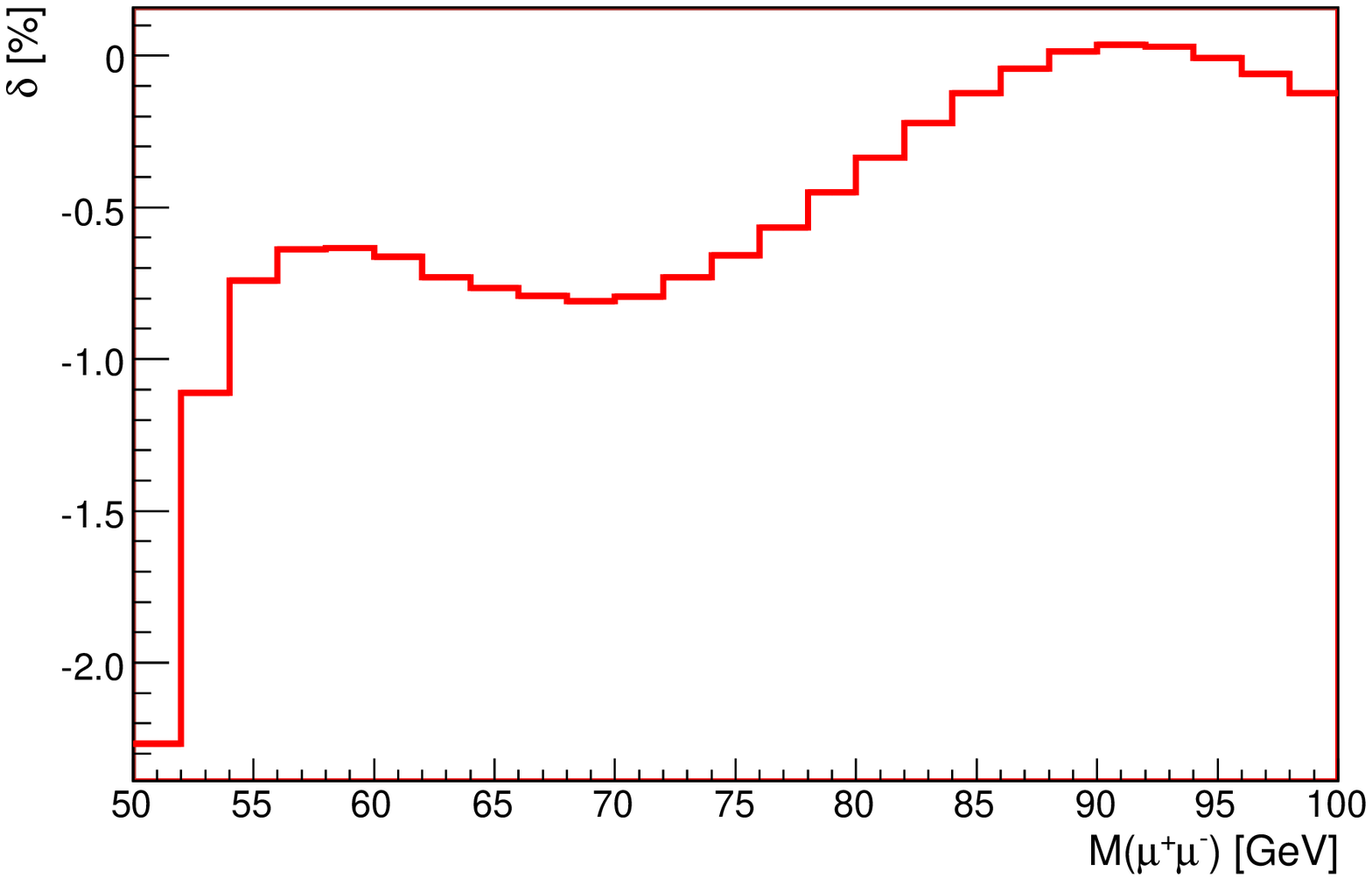} \\
{\bf (a)} \hspace{7cm} {\bf (b)}
\end{center}
\caption{The Born-level NC Drell--Yan cross section and the relative contribution 
of the inverse bremsstrahlung versus the invariant mass of the muon pair.
}
\label{fig5}
\end{figure}
\begin{figure}[h]
\begin{center}
\includegraphics*[width=7.9cm,height=7.6cm,keepaspectratio,angle=0]{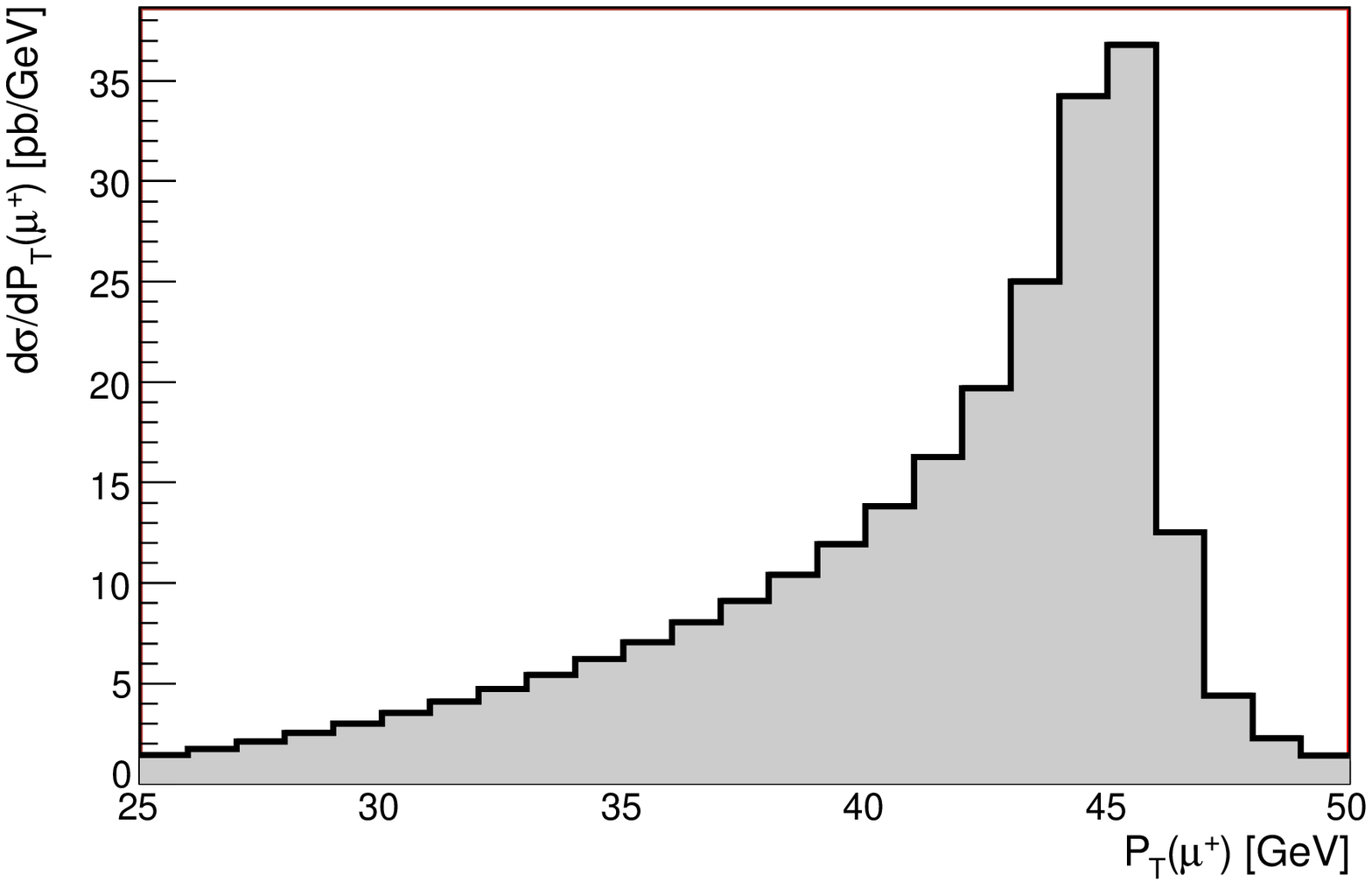}
\includegraphics*[width=7.9cm,height=7.6cm,keepaspectratio,angle=0]{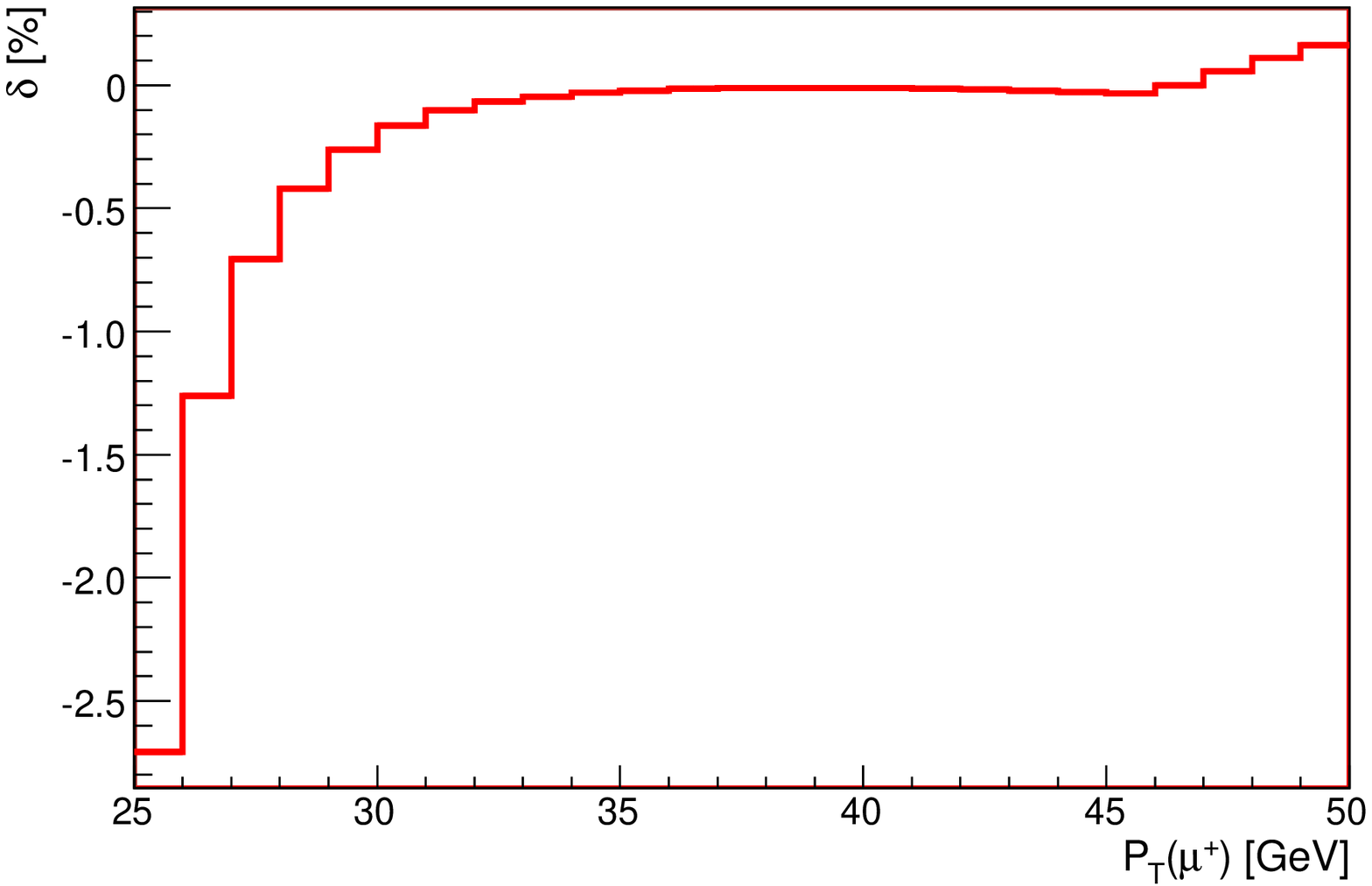} \\
{\bf (a)} \hspace{7cm} {\bf (b)}
\end{center}
\caption{The Born-level NC Drell--Yan cross section and the relative contribution 
of the inverse bremsstrahlung versus the $\mu^+$ transverse momentum.
}
\label{fig6}
\end{figure}

Fig.~\ref{fig5} shows the Born differential cross section of the neutral current Drell--Yan
process ({\bf a}) and the relative correction $\delta_{\gamma q}$ ({\bf b}) as a function of
invariant mass $\mathrm{M_{\mu^+\mu^-}}$ of the muon pair.
Fig.~\ref{fig6} gives us results for the Born differential cross section of the neutral current 
Drell--Yan process ({\bf a}) and the relative correction $\delta_{\gamma q}$ ({\bf b}) 
as a function of $\mathrm{\mu^+}$ transverse momentum $\mathrm{P_{T,\mu}}$.
The distributions around the $W$ and $Z$ resonances are plotted. The drop-offs in the first bins of the
correction distributions in NC have no any physical sense. They arise because the factorization
procedure with the longitudinal partonic density functions doesn't allow to apply the experimental
cuts unambiguously. The drop-offs can be shifted by choosing a different cut value. We checked that 
the rest of the distributions doesn't suffer from this problem.  

In this way we presented the photon-induced contribution to the first order electroweak 
radiative corrections to Drell--Yan processes. For the case of charged current scattering
our results are in a good agreement with earlier calculations of the other group.
The neutral current case was considered in an analogous manner. 
This inverse bremsstrahlung contribution should be taken into account together 
with all other relevant effects to reach the accuracy of the Drell--Yan process 
theoretical description adequate to the precision of the forthcoming LHC experiments. 
The typical size of the contribution is below one percent, but for the case of transverse
momentum distribution in CC scattering, the effect can reach up to 16\% depending on the
cut value. We are going to implement the results of our calculations
into a general Monte Carlo event generator for Drell--Yan processes, which is under
development in the SANC group.


\subsection*{Acknowledgements}

We are grateful to D.~Bardin, S.~Bondarenko, P.~Christova, L.~Kalinovskaya
for fruitful discussions and critical reading of the manuscript.
This work was supported by the RFBR grant 07-02-00932.
One of us (A.A.) thanks also the grant of the President RF
(Scientific Schools 5332.2006) and the INTAS grant 03-51-4007.



\end{document}